\pdfoutput=1

\documentclass[]{spie}  

\usepackage[]{graphicx}
\usepackage{subfigure}
\usepackage{array}
\usepackage{booktabs}
\usepackage{color}
\title{The Cosmology Large Angular Scale Surveyor (CLASS): \mbox{40 GHz optical design} } 

\author{Joseph R. Eimer\supit{a}, Charles L. Bennett\supit{a}, David T. Chuss\supit{b}, Tobias Marriage\supit{a}, \mbox{Edward J. Wollack\supit{b}} and Lingzhen Zeng\supit{a}
\skiplinehalf
\supit{a}Johns Hopkins University, 3400 N. Charles Street, Baltimore, MD 21218 \\
\supit{b}NASA Goddard Space Flight Center
}

\authorinfo{Further author information: (Send correspondence to J.R.E.)\\J.R.E.: E-mail: eimer@pha.jhu.edu}

\begin{document} 
  \maketitle 

\begin{abstract}
The Cosmology Large Angular Scale Surveyor (CLASS) instrument will measure the polarization of the cosmic microwave background at 40, 90, and 150 GHz from Cerro Toco in the Atacama desert of northern Chile.  In this paper, we describe the optical design of the 40 GHz telescope system.  The telescope is a diffraction limited catadioptric design consisting of a front-end Variable-delay Polarization Modulator (VPM), two ambient temperature mirrors, two cryogenic dielectric lenses, thermal blocking filters, and an array of 36 smooth-wall scalar feedhorn antennas.  The feed horns guide the signal to antenna-coupled transition-edge sensor (TES) bolometers.  Polarization diplexing and bandpass definition are handled on the same microchip as the TES.  The feed horn beams are truncated with $10$ dB edge taper by a 4 K Lyot-stop to limit detector loading from stray light and control the edge illumination of the front-end VPM.  The field-of-view is $19^\circ \times 14^\circ$ with a resolution for each beam on the sky of  $1.5^\circ$ FWHM.   
\end{abstract}



\section{INTRODUCTION}
\label{sec:intro}  

The hypothesis that the early universe underwent a period of accelerating expansion, called inflation, has become an essential mechanism for explaining the flatness and homogeneity of the universe and the fluctuations found in the cosmic microwave background (CMB)\cite{Guth, Linde, Albrecht,1982PhRvL..49.1110G,1983PhRvD..28..679B}.  
Inflation predicts the existence of primordial gravitational waves that would have produced a unique polarization pattern called the B-modes on the CMB\cite{1997PhRvL..78.2054S,1997PhRvD..55.7368K}.  Measurement of the amplitude of these gravitational waves can be used to infer the energy scale of the potential driving the expansion\cite{2009AIPC.1141...10B}. Detection of this signal would be a dramatic confirmation of the inflation paradigm and significantly tighten constraints on inflationary models. 

The Cosmology Large Angular Scale Surveyor (CLASS) is a new ground-based instrument designed to search for the large angular scale B-mode signal of the CMB from the Atacama Desert in northern Chile.  This instrument will consist of four separate telescopes: one observing at 40 GHz, two observing at 90 GHz and one observing at 150 GHz. Each bandpass is optimized to observe the CMB through atmospheric windows of high transmission.  Foreground removal, not sensitivity, is expected to limit the detection of the B-mode signal\cite{CMBforegrounds}.  Making observations at multiple frequencies will help to distinguish between the CMB signal and the spectrally distinct synchrotron and polarized dust emission foreground signals. 

Each telescope will employ a Variable-delay Polarization Modulator (VPM) as a fast front-end polarization modulator. This technique will allow the instrument to measure the polarization signal of over $65\%$ of the sky from the Atacama, and could be moved to the Northern hemisphere to observe more sky if appropriate. By targeting the B-mode signal on scales larger than $10^\circ$, CLASS will be able to detect the same B-mode amplitude as experiments with many more detectors but focused on the fainter small scale fluctuations. 

A full description of the CLASS instrument and its overall science objectives will be given in a future paper. In this paper, we focus on the optical design for the 40 GHz telescope.  In Section \ref{sec:overview}, we give a description of the telescope and the procedure followed to generate the design.  Section \ref{sec:analysis} analyzes the performance of the design.  An overview of the construction plans for the optical components is described in Section  \ref{sec:construction}. Stray light and detector loading considerations from spill are discussed in Section \ref{sec:stray_light}.  Section \ref{sec:tol} describes the tolerances for the design.  Finally, Section \ref{sec:conclusions} presents conclusions of this work.

\section{Design Overview} 
\label{sec:overview}

The CLASS 40 GHz telescope will measure the polarization of the CMB from 33 to 43 GHz with 36 transition edge sensor (TES) detector pairs. Each detector pair is coupled through microstrip waveguide (MSWG) to a planar orthogonal mode transducer (OMT) that defines the polarization basis of the detector.  The passband definition, required for bolometric detectors, is performed on chip by MSWG filters between the TES and OMT. Each TES pair, the band defining filters, and the associated OMT are lithographically fabricated on a detector chip. Further details of the CLASS detectors are discussed in Rostem et al.\cite{detectors} In the time reversed sense, modes launched from the OMT pass through a smooth-walled horn antenna and through the rest of the telescope system to the sky.

Independent of the detector properties, the science goal of detecting the faint large angular scale B-mode signal translates into the following requirements for the optical design:
\begin{enumerate}
\item The telescope must have a fast polarization modulator to distinguish the cosmic signal from telescope and atmospheric drifts, i.e. $1/f$-noise. 
\item The modulator must be the first optical component in the telescope to prevent instrument polarization systematic errors.
\item The resolution should be near $1.5^\circ$ to resolve polarization fluctuations greater than $10^\circ$ and to characterize foregrounds on scales smaller than the targeted B-mode signal.
\end{enumerate}
We use horn antennas as the primary beam forming apparatus.  The telescope then redirects and magnifies the symmetric beams formed by the horns. The detector spacing is 38 mm; this is a practical limit to allow enough room for detector mounting and readout circuitry. The telescope design, therefore, is constrained by its two ends - the first optical component will be a polarization modulator and the focal plane will use horn antennas separated by 38 mm as the primary beam forming components.  
 
The immediate consequence of the resolution requirement is that the entrance pupil must be  $\sim37$ cm in diameter. Placing a VPM near this entrance pupil satisfies the fast front-end modulator requirements - further details of the modulator are included in section \ref{sec:VPM}.

With the detector spacing and the entrance pupil specified, the only remaining telescope parameter is the focal length, or equivalently the focal ratio $f/$.  Faster $f/$ choices would increase the field of view (FOV) making it difficult to avoid blockage. Furthermore, the differential light path length through the VPM depends upon the angle of incidence on the device.  The VPM efficiency can be optimized more successfully for a smaller spread of angles.  Slower systems, on the other hand, make the cryogenic system exceptionally long.  While there was no strict optimization, an $f/2$ system is a practical compromise between cryogenic system size and workable FOV.

Since the target scale ($10^\circ$ and larger) is larger than the beam, the mapping speed of the telescope increases as the beam separation on the sky increases and as the spill efficiency increases. For this design, the beam separation is set by the 2.4 $f/ \cdot \lambda$ horn spacing, and the beam truncation is dominated by the cold aperture stop with 10 dB edge taper. 

\begin{figure}[t]
   \begin{center}
  \includegraphics[width=5in]{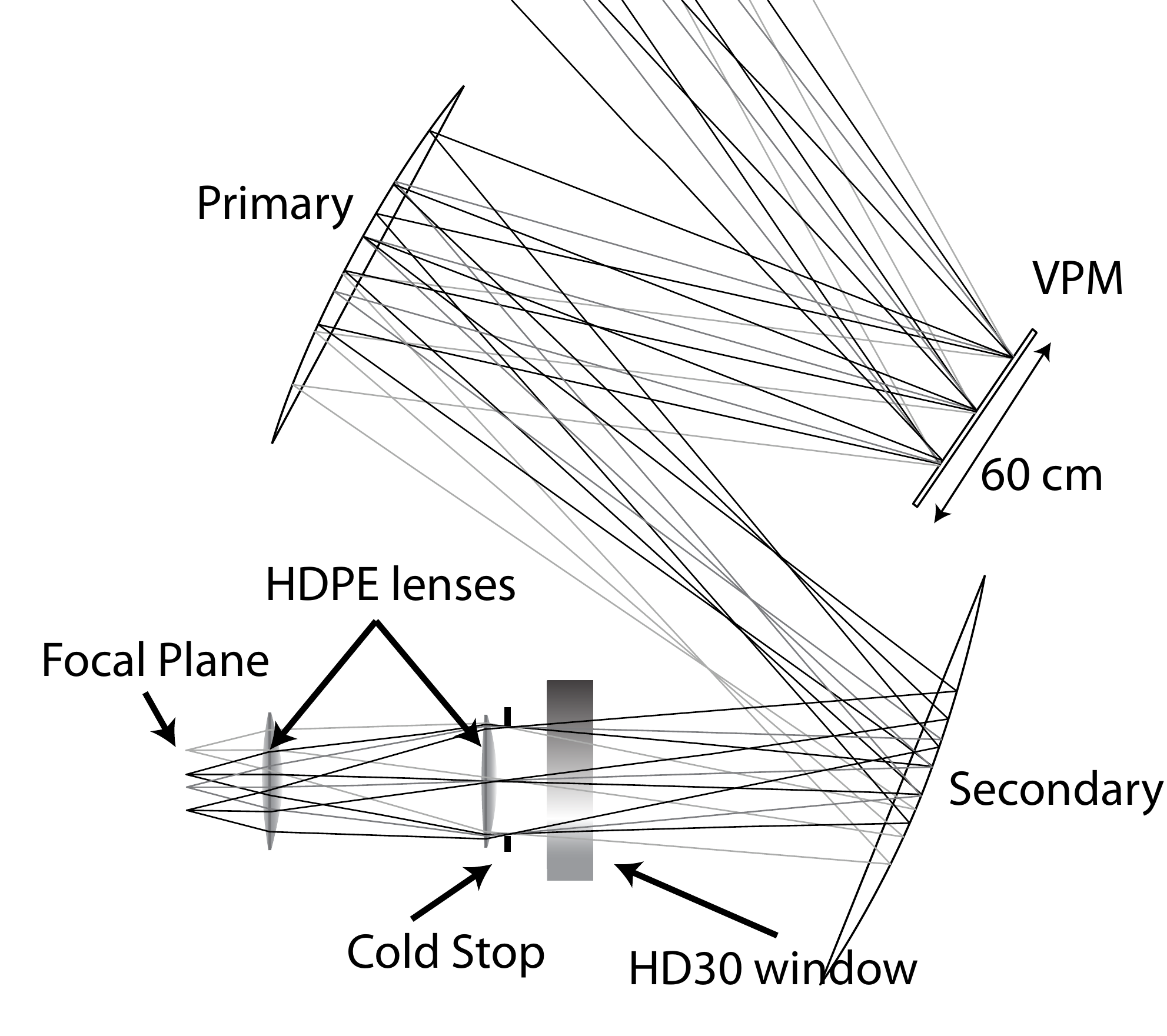}
     \end{center}
   \caption[example] 
   { \label{fig:raytrace} A ray trace of the 40 GHz CLASS telescope is shown for a selection of fields. The first optical element is the VPM. For the purpose of optimizing the telescope design, the VPM is modeled as a flat mirror.  After the polarization of the sky signal has been encoded by the VPM, the light is routed by two mirrors into a cryogenic receiver.  Two cold high-density polyethylene lenses form the final image on the focal plane.}
   \end{figure}

Satisfying the above design criteria required the development of an entirely new type of telescope - the design of which is discussed in the following sections. A ray trace of the final design is shown in Figure \ref{fig:raytrace}. The optical design was generated and optimized using the ray tracing software ZEMAX\footnote{www.radiantzemax.com}. The first element, illustrated as a plane mirror in the Figure, is the VPM.  Following the VPM, the primary and secondary mirrors route the signal through a Zotefoam\footnote{The window will consist of laminated sheets of HD30 manufactured by Zotefoams PLC, www.zotefoams.com} window into a cryostat where high density polyethylene (HDPE) lenses focus the signal onto the horn array. Infrared (IR) blocking filters are situated between the window and the first lens to allow the lenses to cool. The first lens will be cooled to 4 K and the second will be cooled to 1 K.   Parameters summarizing the telescope are in Table \ref{tab:telescope_summary}.

\begin{table}[b]
\caption{\label{tab:telescope_summary}Summary parameters for the CLASS 40 GHz telescope}
\begin{center}
\begin{tabular}{p{2.5in}c}
\toprule
 Beam FWHM & $1.5^\circ$ \\
 Edge taper of cold aperture stop & $\sim10$ dB \\
  $f/$ at 10 dB & 2 \\
 Number of detector pairs & 36 \\
 Entrance Pupil diameter & $\sim 37$ cm \\
 Horn spacing in the focal plane & 2.4 $f/ \cdot \lambda$ at 38 GHz \\
 FOV & $14^\circ \times 19^\circ$ \\
 \bottomrule
\end{tabular}
\end{center}
\end{table}%

\subsection{The Variable-delay Polarization Modulator}
\label{sec:VPM}

CMB polarization instruments must modulate the incoming light to distinguish the cosmic signal from detector, instrument, and atmospheric drifts, e.g. $1/f$ noise.  The modulation strategies employed by most other CMB telescopes involve fast azimuth scanning or rotating a wave plate. Azimuth scanning, unfortunately, modulates both the CMB polarization and the CMB temperature anisotropy synchronously. An experiment adopting this strategy must employ additional techniques to guard against systematic errors mixing these signals. Rapidly rotating a half-wave plate overcomes this limitation but adds the complication of needing to understand the detailed spectral transmission of the wave plate including any anti-reflection (AR) coatings. Modulated thermal emission must also be accounted for in warm wave plates, and fast rotating cryogenic wave plates have proven extremely difficult to deploy. Furthermore there are practical limits to the diameter of wave plates that can be manufactured. This size constraint limits how close to the sky the wave plate can be placed in the optical train, especially for low frequencies. 

CLASS will utilize the new VPM technology as a front-end modulator to distinguish the cosmic signal from various noise sources while circumventing the challenges faced by other techniques. This modulator consists of a polarizing wire array backed by a parallel movable mirror. The operating principle of a VPM has been discussed previously\cite{vpm_theory}, and an example of a VPM fielded in the Hertz instrument is discussed in Krejny et al.\cite{vpm_example}. A few aspects of VPMs that guide the design of this optical system are highlighted here.
\begin{description}
\item[The VPM is the first optical component.] The scalable mechanical construction of a VPM enables the device to be placed at the entrance pupil of a modest-to-small size telescope. Since the resolution required to target large angular scales is low, a VPM with an aperture diameter of 60 cm is sufficient.
\item[The VPM is the polarization modulator.] Each modulation cycle of the VPM is an \emph{absolute} measurement of a Stokes parameter in a particular direction. No scan modulation is required. 
Since data can be collected while drift scanning or slow azimuth scanning, the tolerance analysis need not include telescope bending modes that would otherwise be present during scan turnarounds. 
\item[The VPM transmission spectrum is simple.] The VPM is a purely reflective device. The lack of dielectric substances make the spectral characteristics of the VPM simple to understand.  Chuss et al. \cite{2012ApOpt..51..197C} have identified resonant conditions that exist for particular VPM arrangements, but these situations are well understood and can be included in the transfer function template of the VPM. 
\end{description}

\begin{figure}[t]
   \begin{center}
  \subfigure[]{\label{fig:VPMcartoon}
  \includegraphics[width=0.35\textwidth]{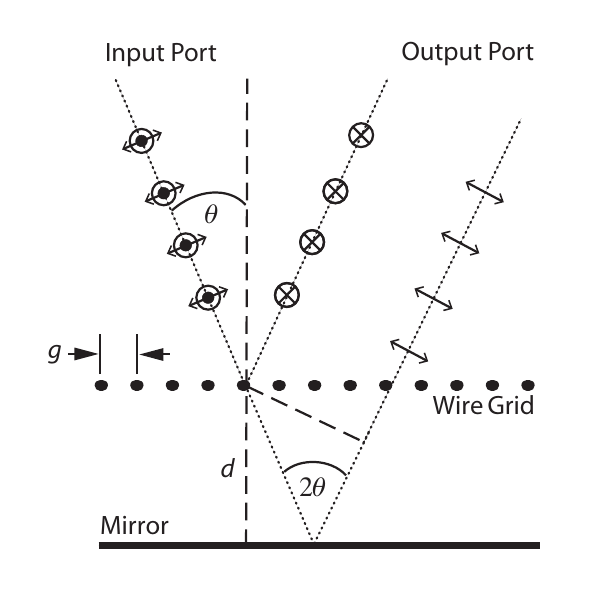}}
  \subfigure[]{\label{fig:Poincare}
  \includegraphics[width=0.35\textwidth]{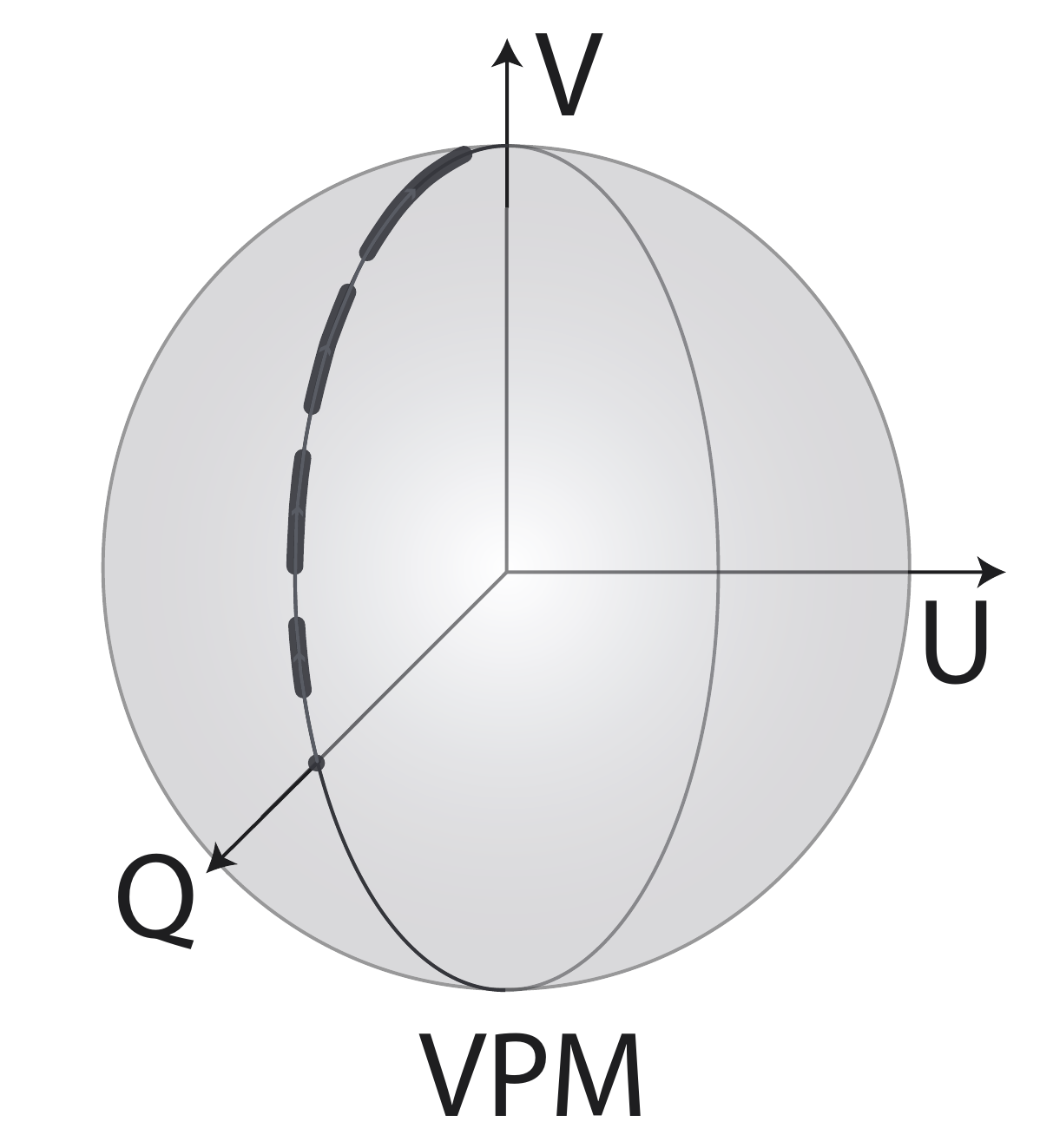}}
   \end{center}
   \caption[example] 
   { \label{fig:VPM} (a) A schematic of the VPM operation is shown. The input sky signal is separated by the polarizing grid into two orthogonal components. One component is reflected from the wires, and the other component passes through the wires and is reflected by the movable mirror.  The two reflected signals combine to form the modulated output signal.  (b) The dashed path on the Poincar\'{e} sphere illustrates the change in sensitivity to the sky basis defined Stokes parameters. As the mirror moves, the measured parameter modulates between Stokes Q and V.}
   \end{figure}

Figure \ref{fig:VPMcartoon} shows a schematic of the VPM layout. An incoming signal is separated into its two orthogonal linear polarizations by the wire array.  The component polarized parallel to the wire array is reflected, while the component with the polarization orthogonal to the wires passes unimpeded. The transmitted component then reflects off the movable mirror and passes back through the wire array. The relative phase between the two reflected components is specified by controlling the separation between the wires and the mirror. 

The signal for a given direction on the sky can be represented by Stokes parameters $I_{sky}$, $Q_{sky}$, $U_{sky}$ and $V_{sky}$.  The orthogonal antenna pair of the detector projected through the optics onto the sky naturally defines $Q_{det}$ through detector differencing. When the detector basis projected onto the VPM is rotated $45^\circ$ with respect to the VPM wires, the polarization sensitivity of the detector to the sky signal as a function of the relative phased delay, $\phi$, is
 \[Q_{det} = Q_{sky} \cos \phi + V_{sky} \sin \phi. \]
 Figure \ref{fig:Poincare} shows a portion of this modulation cycle on the Poincar\'{e} sphere. 
 
 Periodically the telescope will execute $45^\circ$ rotations about the telescope boresight.   Such a rotation changes the detector sensitivity relative to the sky-fixed Stokes basis to 
  \[Q_{det} = U_{sky} \cos \phi + V_{sky} \sin \phi .\]
In this way CLASS will have complete coverage of the Stokes parameters for every point within the accessible sky region. The $V_{sky}$ astrophysical signal is expected to be zero and will serve as critical null observations for systematic error checking. 
\subsection{Fore-optics design and description}
\label{sec:fore_optics}

The telescope fore-optics consist of the VPM and the two curved mirrors shown in Figure \ref{fig:raytrace}. The two curved mirrors create an image of the cold 4 K aperture stop on the central region of the VPM.  In this location, the diameter of the VPM is minimized for a fixed entrance pupil size. There is no magnification requirement of this relay system for this low resolution telescope. Note the entrance pupil is purposefully tilted with respect to the VPM to appear circular when viewed from the center of the FOV.

\begin{figure}[t]
   \begin{center}
  \includegraphics[width=6.5in]{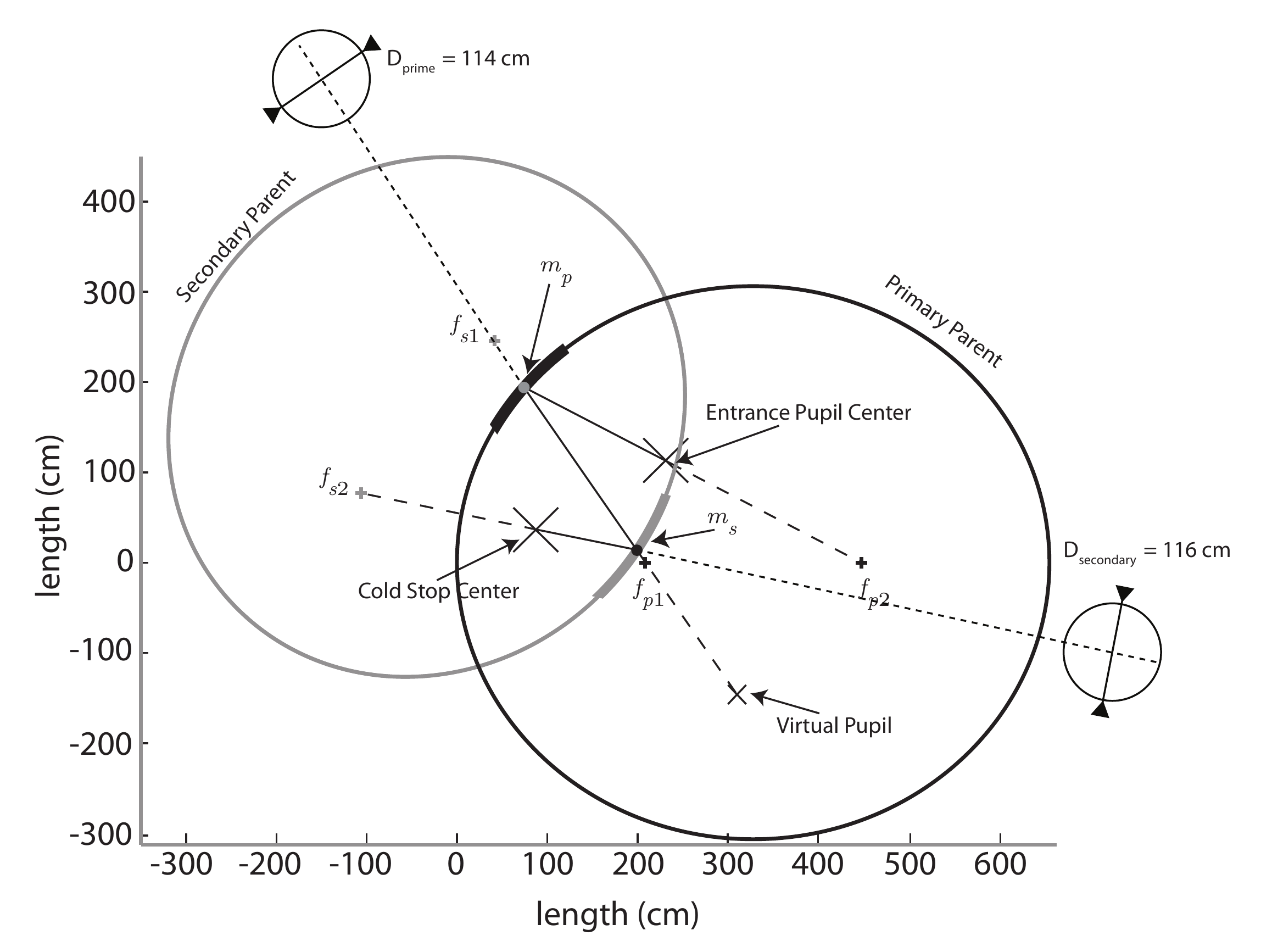}
   \end{center}
   \caption[example] 
   { \label{fig:foreoptics} The relative geometry of the fore-optics mirrors is shown. Each ellipse is rotated about its major axis to generate the surface of the mirrors.  The portion of the ellipsoids used for the mirrors are indicated by thicker lines on the ellipses. The darker ellipse generates the primary. Crosses are placed at the focal points of the ellipses.  The large `X' mark indicates the location of the cold stop and the entrance pupil.  The smaller `x' indicates the intermediate virtual pupil.  Both mirrors have circular perimeters when projected along the central ray light path.  The dotted lines indicate this projection direction for each mirror.  The projected diameters of the primary and secondary mirrors are 114 cm and 116 cm respectively.  }
   \end{figure} 

The fore-optics were optimized as an afocal system. It was important during this step to temporarily place the stop surface in ZEMAX at the VPM.  With the stop as the first optical surface, ZEMAX efficiently launches all necessary rays for the optimization.  If the stop surface was kept in its physical location, the aperture stop vignetting function would depend upon the exact size, shape, and location of all prior optical elements. Optimization for this off-axis tilted system would not be computationally feasible. In the temporary `false stop' arrangement, the goal of the fore-optics system is to map plane waves, apodized by the stop, to aberration-free spherical waves at the physical location of the stop. The location and shape of the two mirrors were allowed to vary in order to optimize the following criteria:
\begin{itemize}
	\item Minimize the wave front error (WFE) of the outgoing spherical waves for each incoming plane wave direction.
	\item Ensure marginal rays were sufficiently far from mirror edges and the cryostat edges to prevent aperture blockage.
	\item Minimize the distance between the intersection of the chief rays for all sampled directions at the cold stop.
	\item Minimize the distance between the top marginal rays for all fields and a point 15 cm from the chief ray center in the desired pupil plane. 
	\item Ensure the distance between the two powered mirrors was not larger than 2.2 meters (to limit the size of the telescope).
	\item Ensure the intended aperture stop plane is always orthogonal to the chief ray of the central field.
\end{itemize}
 A damped least squares (DLS) optimization resulted in an initial design, but the ZEMAX `global search' optimization was used to generate the final design. Once the optimization converged to an acceptable solution, the `false stop' surface in ZEMAX was removed and a stop with 30 cm diameter was placed in the actual aperture stop location. 

Single off-axis mirror designs for the fore-optics were ruled out because of unavoidable polarization distortion\cite{1987IJIMW...8.1165M}. In addition to correcting this distortion, the degrees of freedom added by using a second mirror enable an improved entrance pupil. Figure \ref{fig:foreoptics} shows the relationship between the two fore-optics mirror shapes resulting from the optimization. Both mirrors are off-axis sections of ellipsoids generated by rotating the parent ellipse shape about it major axis. The mirrors are conveniently defined by the focal points of the parent ellipse and a point at the center of each mirror. These coordinates are given in Table \ref{tab:mirror}. This virtual pupil is then imaged onto the VPM by the primary mirror.  The location of these points are indicated by `X' marks in the Figure. 
 
\begin{table}[t]
\begin{center}
\caption{\label{tab:mirror}Parameters defining the shape and relative geometry of the fore-optics. The coordinate system origin is  the vertex of the primary mirror ellipsoid.}
\begin{tabular}{lll}
 Point & x (cm) & y (cm)\\
 \toprule
 $f_{p1}$ & 208.543348 & 0\\
 $f_{p2}$ & 446.483014 & 0 \\
$ f_{s1}$ & 40.759849 & 245.895577 \\
 $f_{s2}$ & -106.455190 & 76.696615 \\
 $m_p$ & 73.410690 & 192.517211\\
$ m_s$ & 11.566104 & 198.577111\\
 Stop Center & 83.430267 & 37.653512 \\
 \bottomrule
\end{tabular}
\end{center}
\end{table}

\subsection{Re-imaging optics}
\label{sec:reimaging_optics}

Light from a specified field passing through the cold stop is nearly a spherical wave. The re-imaging optical system is therefore more concerned with re-imaging these nearly-spherical waves to a flat focal surface than correcting aberrations.

The lenses for the re-imaging optics are made of HDPE. A survey of the literature suggests the room temperature index of refraction for HPDE is near $n=1.526$ at 40 GHz\cite{2006ApOpt..45.5168M,afsar85}. Since the linear dimensions of HDPE contract by $2\%$ when cooled from 300 K to 4 K\cite{HDPE_contraction}, we used the Lorentz-Lorenz relation\cite{Jackson} 
\[ \rho \propto \frac{n^2 -1}{n^2+2}\] 
to scale the room temperature index to $n=1.564$ at 4 K and 1 K. This cold index value is assumed for the re-imaging optics design. 

Designs using a single HDPE lens were found to have diffraction limited performance across the required FOV, but the lens for such a design was too large to be practically cooled. A design using two smaller and thinner lenses was adopted instead. The symmetric signal through the stop enables both lenses and the focal plane to be co-axial with the aperture stop.

Several criteria were used to optimize the re-imaging optics
\begin{itemize}
	\item The WFE for each field at the focal plane was minimized. 
	\item The focal ratio ($f/$) was optimized to be near 2.
	\item Moderate clearances between the pupil lens and the aperture stop, between the two lenses, and between the final lens and the focal plane were enforced.
	\item All field chief rays should intersect the focal plane as close as possible to $90^\circ$ 
	\item Each lens must be thinner that 5.4 cm with a minimum edge thickness of 1.1 cm.
\end{itemize}
Similar to the fore-optics, a DLS optimization produced an initial design, and a global search produced the final design. Both lens for the final design are convex-convex with on-axis ellipsoidal surfaces.  Each surface is conveniently described through the sag equation
\[ z = \frac{ r^2/R}{1+ \sqrt{1-(1+\kappa)r^2/R^2}}\]
where $z$ is the distance from the vertex tangent plane to the lens surface, $r$ is the radial coordinate off the lens axis, $R$ is the radius of curvature at the vertex of the lens, and $\kappa$ is the conic constant. Table \ref{tab:lens} gives the shape parameters and location of each lens.  Note the AR coating thicknesses are not included in the separation distances provided. 

\begin{table}[t]
\begin{center}
\caption{\label{tab:lens} Geometric parameters for the re-imaging optics. AR coating thicknesses are not included in the distance and thickness parameters.}
\begin{tabular}{lccc}
\toprule
\multicolumn{4}{l}{Distance between the stop and the first lens = 2.999992}\\
\midrule
 \multicolumn{4}{c}{\textbf{First Lens parameters at 4 K}}\\
\midrule
 Side & R (cm) & $\kappa$ & central thickness (cm)\\
 \midrule
 1 & 152.420594 & --3.790630 & 4.659434 \\
 2 & 53.948154 & -1.705192 & \\
 \midrule
 \multicolumn{4}{l}{Distance between lenses = 56.151877}\\
 \midrule
\multicolumn{4}{c}{ \textbf{Second Lens parameters at 1 K}}\\
\midrule
 Side & R (cm) & $\kappa$ & central thickness (cm)\\
 \midrule
 1 & 122.103054 & -22.013666 & 5.4 \\
 2 & 49.471025 & -2.654538 & \\
 \midrule
 \multicolumn{4}{l}{Distance between the second lens and the focal plane = 21.998085}\\
 \bottomrule
\end{tabular}
\end{center}
\end{table}

\subsection{Focal plane}
\label{sec:focal_plane}
\begin{figure}[t]
   \begin{center}
  \subfigure[]{\label{fig:FPlayout}
  \includegraphics[width=2in]{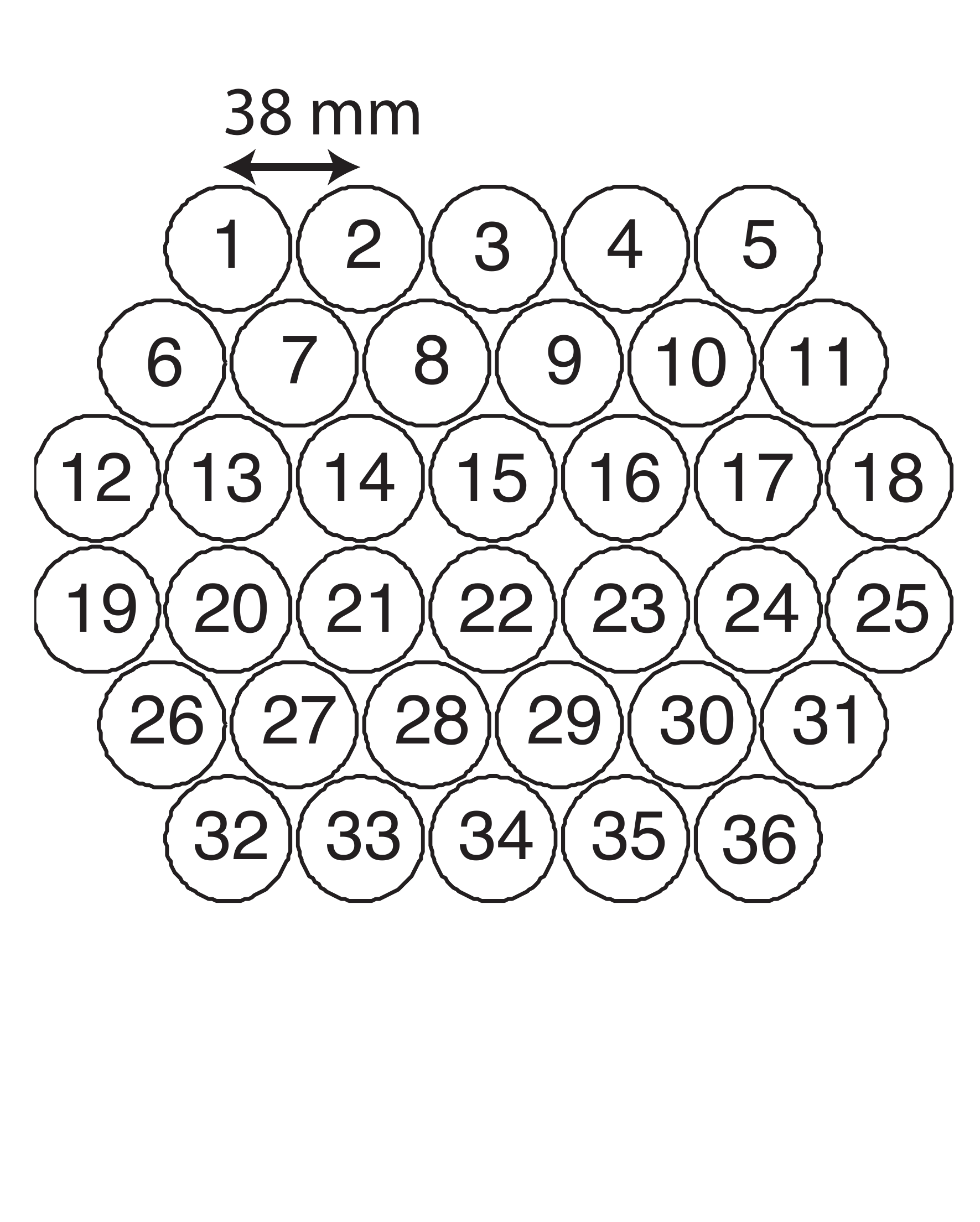}}
  \subfigure[]{\label{fig:profile}
  \includegraphics[width=1.4in]{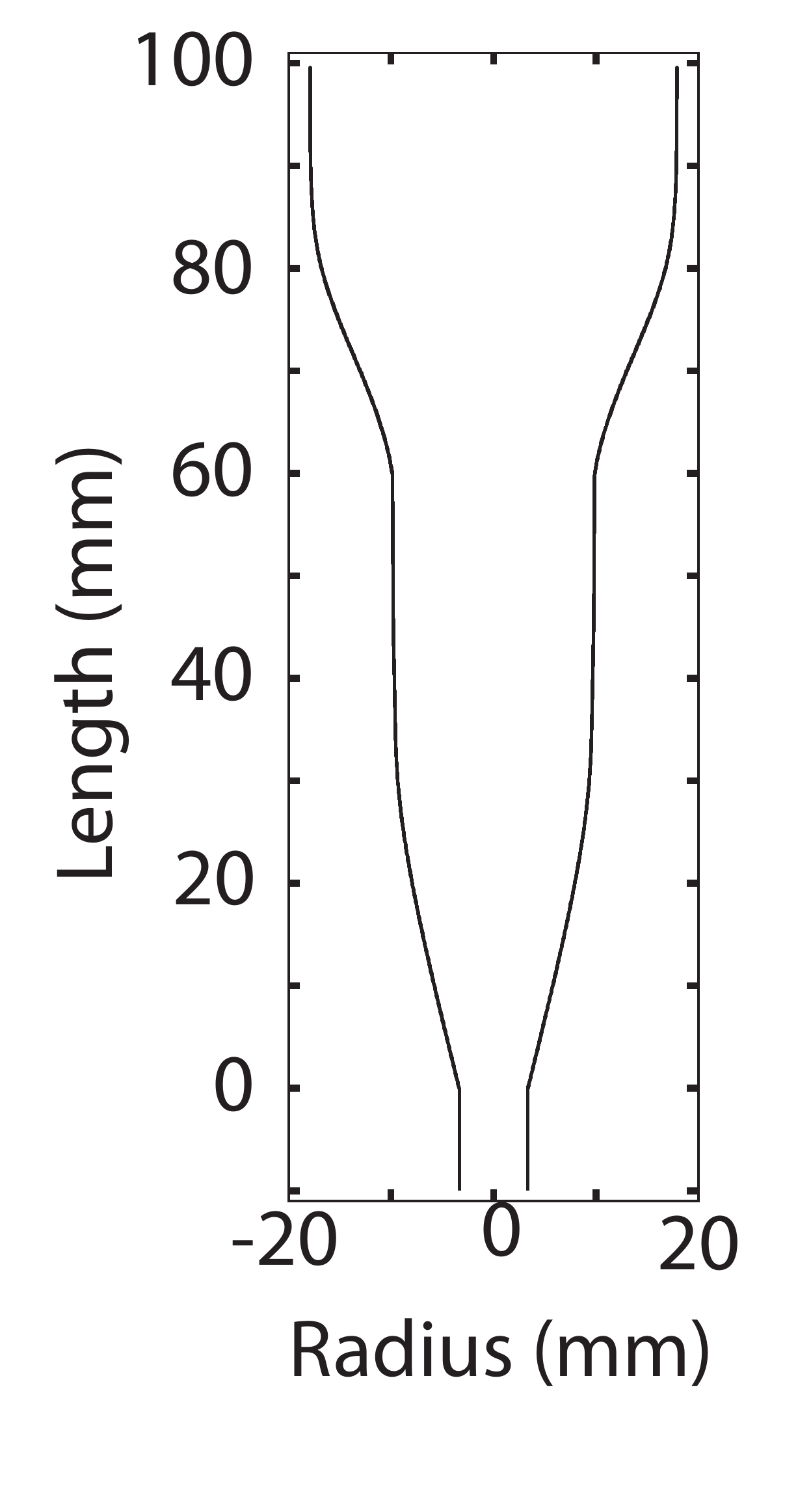}}
   \subfigure[]{\label{fig:hornbeam}
   \includegraphics[width=3.1in]{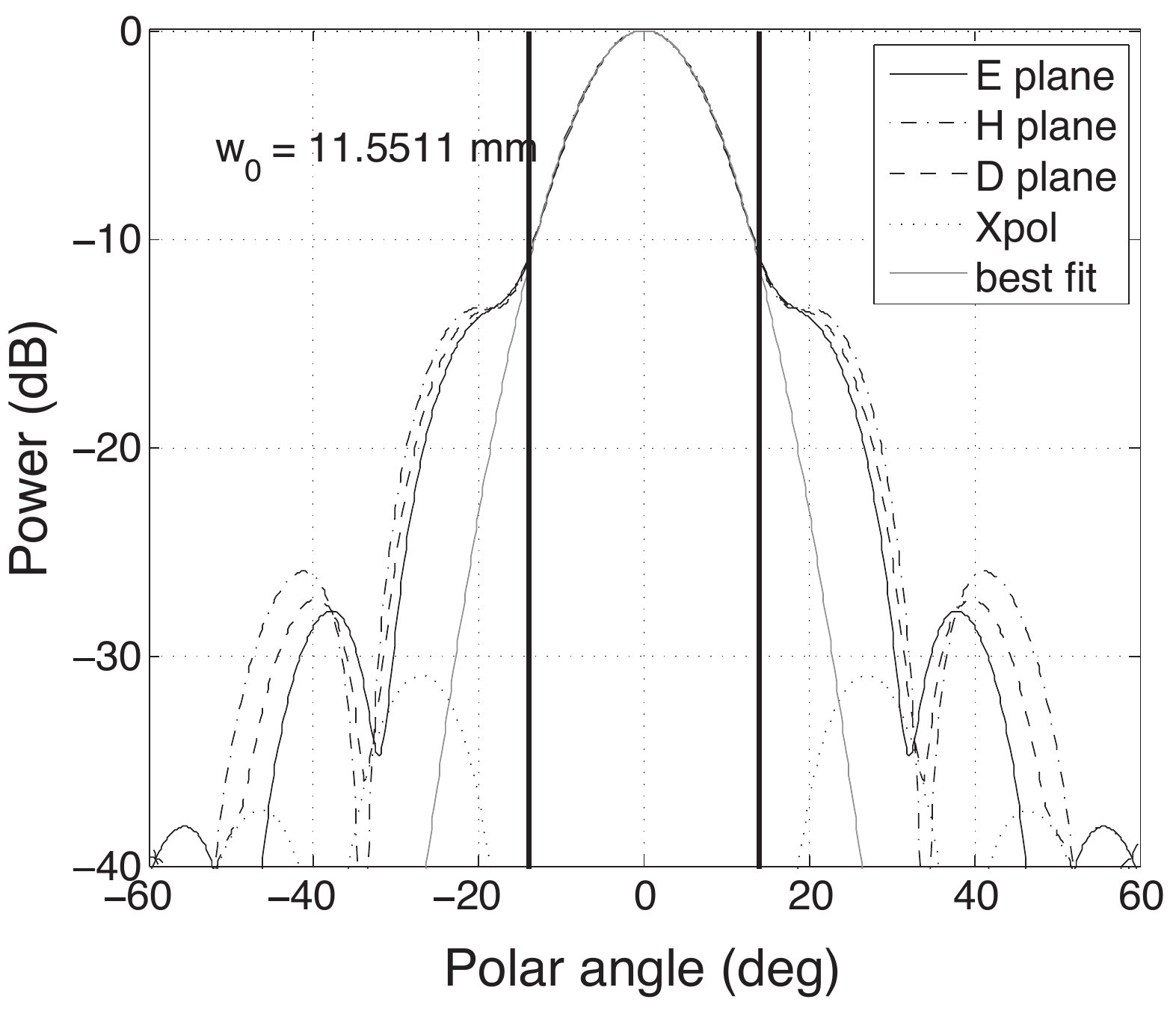}}
   \end{center}
   \caption[example] 
   { \label{fig:fp} (a) The arrangement of the horns in the focal plane is shown. The horns are arranged in two hex-close-packed halves.  The nearest-neighbor horns are separated by 38 mm. (b) The horn profile of the 40 GHz smooth walled horn is shown. The length of the contoured section of the horn is 100 mm. The horn aperture inner diameter is $\sim 36$ mm. A short section of circular waveguide is include in the Figure. (c) The predicted far field beam pattern of the horn is shown along with the best fit Gaussian beam to the region between the vertical lines. The vertical lines are at $\pm 14^\circ$ and show the angle represented by marginal rays in the ray trace.}
   \end{figure}

The telescope ultimately couples the signal from the sky to a flat focal plane.  The focal plane uses 36 smooth-walled horn antennas to guide the incoming radiation to the CLASS detectors. Each horn will be smooth-walled with a monotonic profile\cite{LZthesis}. While this architecture is more difficult to design than corrugated horns, the smooth interior surface of the horn is easier and cheaper to machine and competitive performance is achievable.  The horn design procedure for this telescope is described by Zeng et al.\cite{2010ITAP...58.1383Z}. To be compatible with the telescope, the horns have been designed to produce a 10 dB edge taper at $f/2$. This constraint, combined with minimizing return-loss and cross-polar\footnote{We exclusively use Ludwig's third definition\cite{Xpol} of cross-pol in this paper.}  response, form the basic constraints for the horn optimization. The horn/telescope system design is further described in section \ref{sec:analysis}. The horn arrangement in the focal plane and the horn shape are shown in Figures \ref{fig:FPlayout} and \ref{fig:profile} respectively.

\section{Performance analysis}
\label{sec:analysis}

\begin{figure}
   \begin{center}
  \includegraphics[width=4in]{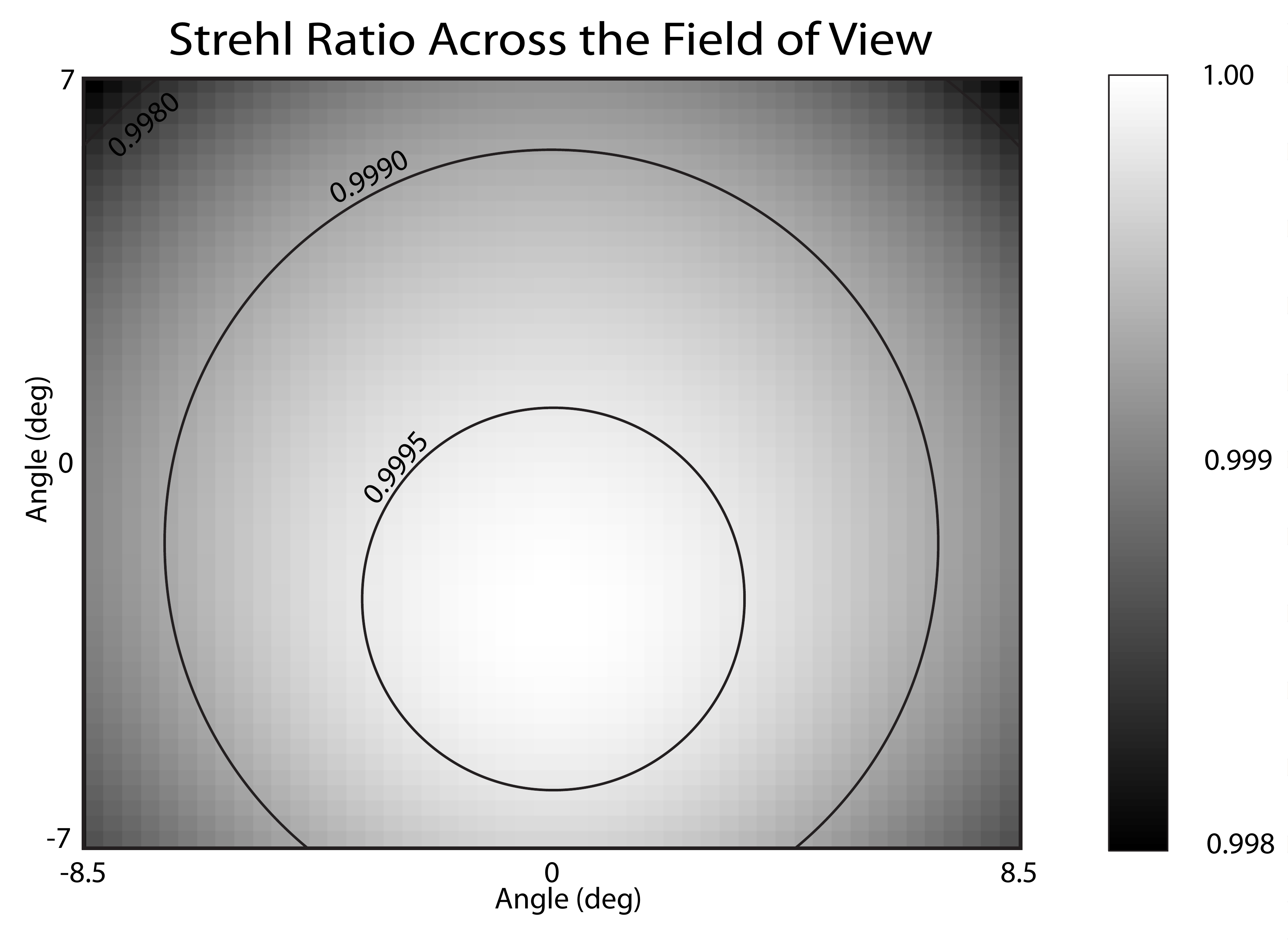}
   \end{center}
   \caption[example] 
   { \label{fig:StrehlRatio} The Strehl ratio for all points within the FOV are displayed.  The Strehl Ratio is above 0.998 for all field points within the FOV.}
   \end{figure}

Due to the speed rays can be traced by a computer through the optical system, the ray-calculated WFE metric was used for optimizing the location and shape for all optical surfaces. The maximum rms WFE across the FOV is $8.74 \times 10^{-3}$ waves for the final design.  The rms WFE is frequently used by ray tracing software to approximate the Strehl ratio, S, of the telescope via the equation
\[S = e^{-(2 \pi \sigma)^2} \]
where $\sigma$ is the rms WFE referenced to the geometric ray centroid at the focal plane. 
This approximation is accurate when the peak of the point spread function (PSF) is well defined and the Strehl ratio is not far \mbox{from 1}. By convention Strehl ratios above 0.8 are considered diffraction limited. From Figure \ref{fig:StrehlRatio}, we see the Strehl ratios for this telescope are  $>0.996$ for all points within the FOV. 

While the Strehl ratio is a useful metric to understand the image quality of a telescope, the fact that the aperture stop is only 30 cm (38 waves) in diameter indicates diffraction effects must be included to describe the telescope performance. To calculate the beams produced by the telescope on the sky and to predict the loading on the detectors from diffraction around the optical elements, the beam pattern of the horns in the focal plane must be included in the analysis.  The field distribution at the aperture of the horn is known, but it is convenient to simplify the horn beam pattern to a Gaussian beam fit to the main beam of the far-field pattern of the horn.  This Gaussian beam accurately represents the portion of the beam passing through the cold stop. Figure \ref{fig:hornbeam} shows the far field beam and the best fit Gaussian to the main beam.  The 11.5511 mm diameter waist producing the best fit beam was launched from the focal plane in the location of the pixel being studied.

\begin{figure}
   \begin{center}
  \includegraphics[width=6.8in]{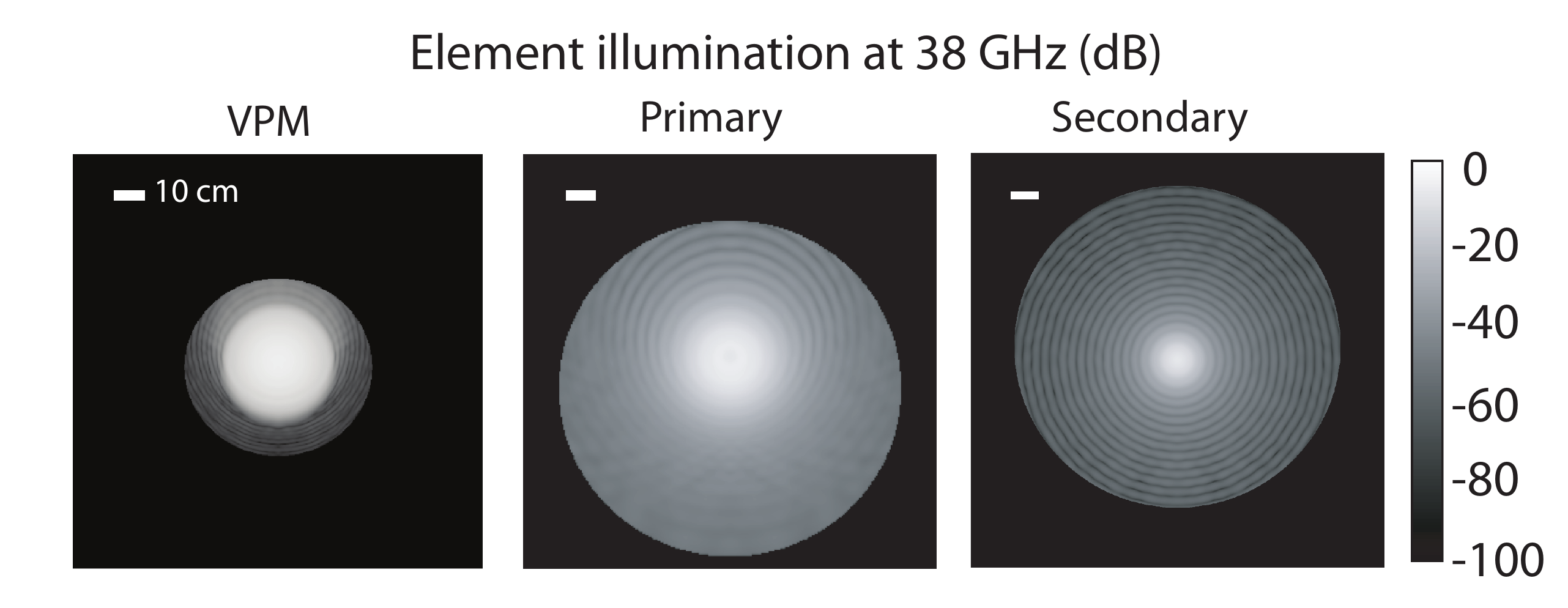}
   \end{center}
   \caption[example] 
   { \label{fig:illum} The illumination of the VPM, primary, and secondary are shown for horn 22. The VPM edges are under illuminated to prevent modulated side lobes. The clear Airy pattern on the secondary is due to the proximity to the intermediate focal surface between the primary and secondary mirrors. }
   \end{figure}

The physical optics propagation (POP) feature in ZEMAX uses scalar diffraction theory to progressively calculate the field distribution on each optical surface. Scalar diffraction theory can not predict the field distribution on aperture edges, but all edges are under-illuminated since the primary beam forming components are the horns themselves.
Beyond the scalar field approximation, ZEMAX also assumes the field component along the direction of travel of the wave is small compared to the transverse field components. This assumption breaks down for very fast systems (typically $\sim f/1$). Since the fastest beams in this telescope are $f/2$, the ZEMAX POP feature is acceptable for understanding the propagation of the fields through the telescope itself. Figure \ref{fig:illum} shows the illumination for a central pixel on the VPM, primary, and secondary. The size of each element was selected to maximize the telescope gain and limit warm spill over the edges. The secondary mirror was enlarged the most because proximity to the intermediate focal surface between the mirrors causes a slow Airy-like radial intensity drop-off and because all spill beyond this element views the \mbox{300 K }environment. Spill over the primary will largely be directed toward the cold sky and therefore does not require as aggressive oversizing.

\begin{figure}
   \begin{center}
  \includegraphics[width=6.5in]{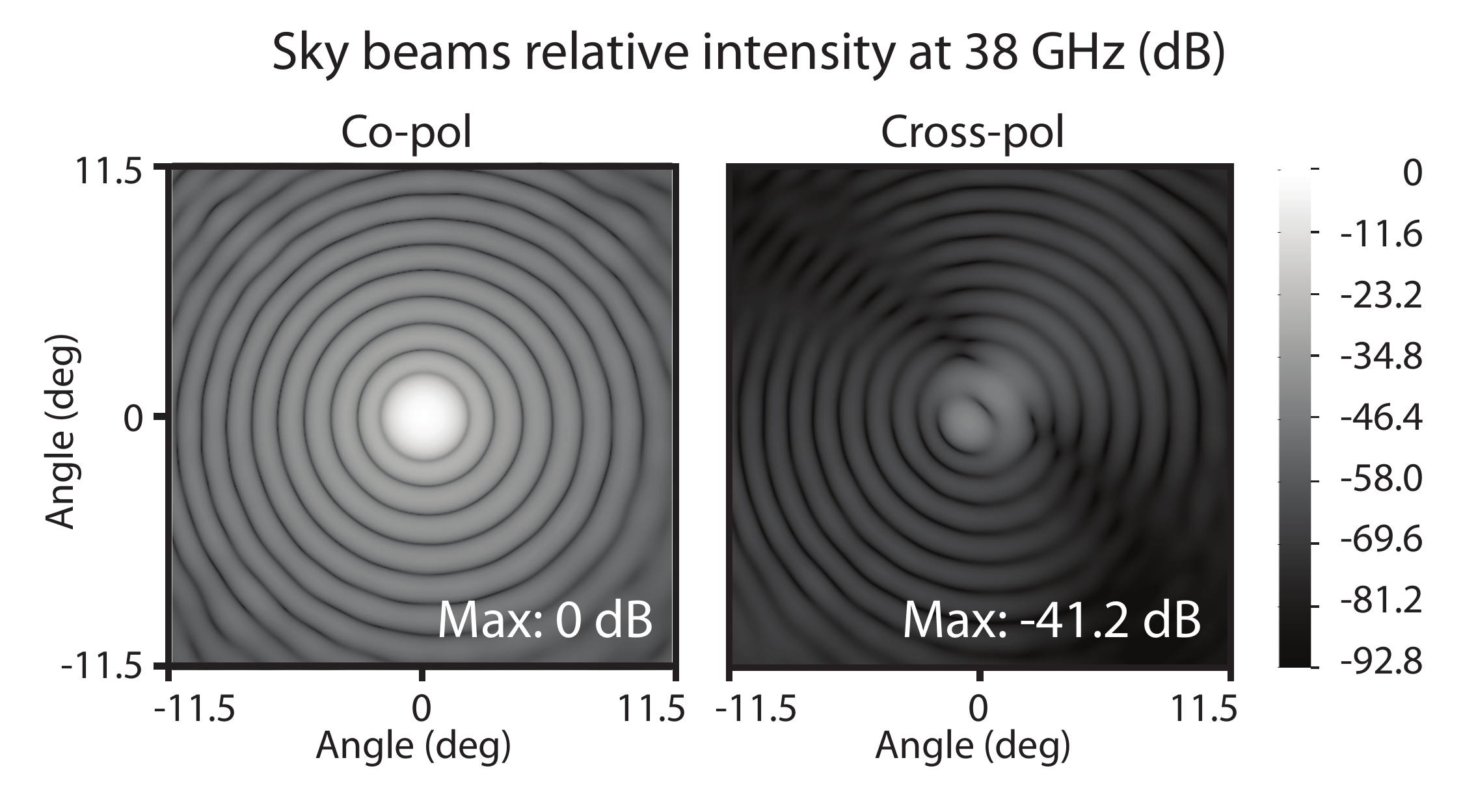}
   \end{center}
   \caption[example] 
   { \label{fig:beams} The GRASP10 calculated co-polar (left) and cross-polar (right) beams for the telescope are shown for a central pixel. The peaks of the cross-polar pattern are more than 40 dB lower than the co-polar peak. The beams for other pixels are similar. The $45^\circ$ rotation of the detector with respect to the plane of symmetry of the telescope is evident in the cross-polar beam pattern.}
   \end{figure}

The final propagation step from the entrance pupil of the telescope to the sky was calculated using GRASP10\footnote{www.ticra.com}.  In GRASP10, the field configuration from ZEMAX is converted to a fictional current distribution that would have produced the input field distribution.  The fictional current distribution is then used to calculate the far-field beam pattern.  Figure \ref{fig:beams} shows the predicted co-polar and cross-polar beams of a typical pixel.  The FWHM of the beams on the sky is $\sim1.5^\circ$. The cross-pol peaks are seen to be more than 40 dB below the main beam peaks. The main beam is circular with the first side lobe showing some asymmetry, but is roughly 23 dB lower than the main beam peak.

\section{construction plan}
\label{sec:construction}

The VPM for the 40 GHz telescope will consist of a planar array of regularly spaced wires backed by a parallel and movable aluminum faced honeycomb mirror. The lightweight, high stiffness properties of the honeycomb material enable us to rapidly translate the mirror while staying parallel to the wire array. A prototype VPM wire array has been constructed with a 50 cm clear aperture\cite{grid}.  The prototype array uses 64 $\mu$m diameter gold plated tungsten wire regularly spaced at 200 $\mu$m pitch.  The minimum resonant frequency of the wires was measured to be above 150 Hz, well above the modulation frequency of the VPM.  To maintain good phase definition, the wires are planar within 1\% of a wave.  The techniques used for the prototype are expected to scale to the required 60 cm clear aperture without significant modification. The polarization transfer function of the prototype has been measured, and the device performs as expected\cite{2011RScI...82h6101E}. 
   
\begin{figure}
   \begin{center}
  \includegraphics[width=5in]{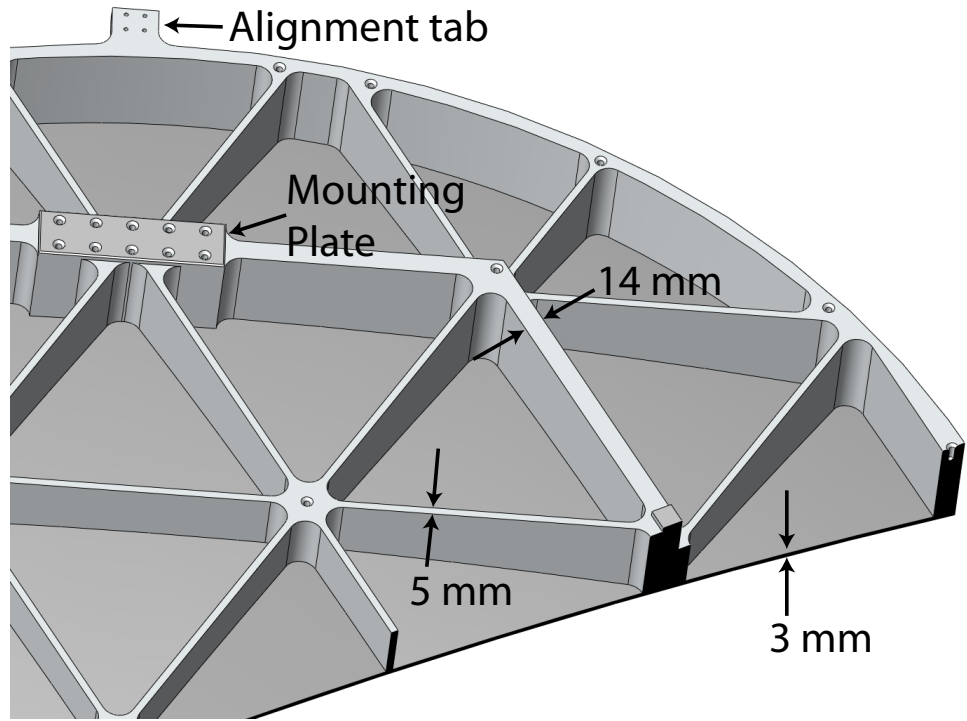}
   \end{center}
   \caption[example] 
   { \label{fig:secondary} A sectioned view of the secondary mirror is shown.  The face sheet of the mirror is 3 mm thick.  5 mm wide webs are arranged in a triangular pattern to stiffen the mirror.  A single 14 mm wide web encircles the mirror and ensures the mirror stiffness is greater than the mounting hardware stiffness. Three mounting plates are arranged in a regular triangular pattern of the the thick web. Tapped holes are placed at webbing intersections to mount thermal monitoring and control hardware.  The primary mirror will have an analogous construction.}
   \end{figure}

The primary and secondary mirrors will both be open-back light-weighted monolithic aluminum structures. Figure \ref{fig:secondary} shows a sectioned view of the secondary mirror. This mirror has a minimum 3 mm face sheet with 5 mm thick webbing arranged in a triangular pattern.  A single 14 mm thick web encircles the back of the mirror for mounting. Standard CNC machining techniques are more than sufficient for achieving the generous tolerances required for this low frequency channel. The mirrors, however, will be constructed such that tolerances limits even for the 90 GHz channel are satisfied.  This will enable the same mechanical design to be iterated for the higher frequency channels and will not significantly increase the difficulty of manufacturing. The surfaces will be hand-finished to a $\sim 2$ $\mu$m rms surface roughness to remove possibly polarizing tool marks. This roughness will degrade the gain by the factor 
\[\alpha = e^{-\left( 4 \pi \epsilon/\lambda \right)^2}\]
where $\epsilon$ is the rms surface roughness \cite{Ruze}.  For an $\epsilon=2$  $\mu$m surface finish, the gain degradation will be negligible for the 40 and 90 GHz bands. At the same time, the surface is still rough enough to avoid specularly reflecting high frequency IR and visible light.  Avoiding specular reflection of higher frequency signals will help prevent unintentionally focusing energy on the cryostat or other portions of the telescope.

HDPE was selected as the lens material for its low dielectric loss, ease of machining, ease of AR coating, and the availability of raw material in sufficiently large pieces to construct the lenses. Each lens will be AR coated by a simulated dielectric layer.  This layer will consist of small holes drilled in a rectangular array across the lens surface. The designed geometric lens parameters will be enlarged to accommodate the  $2\%$ integrated linear expansion when warmed from 4 K to 300 K \cite{HDPE_contraction}. Thermal radiation blocking filters at 60 K and at 4 K will be used to ensure the lenses get cold.  These filters will consist of metal-mesh frequency selective surfaces and neutral density dielectric slabs. 

The aperture stop will be constructed from a formed ring of steel loaded epoxy called steelcast\cite{steelcast}. The ideal cold stop would be an infinitesimally thin aperture allowing all radiation within a defined radius to pass unimpeded and all radiation beyond that radius to be blocked. Unfortunately such a surface does not exist at any wavelength. More accurately, a stop should be expressed as a surface with a particular absorption as a function of incident angle and polarization. To increase the loss, and approach the perfectly absorbing ideal, the thickness of the absorber must be a significant fraction of a wavelength (7 mm). Even in this case, the stop is not perfectly `black.' To decrease the edge illumination on the stop, a series of axillary rings with slightly larger apertures will be placed on both sides of the geometric-optics defined stop.  Each aperture will gradually clip the beam with the central geometrically located ring dominating the clipping. Sharp et al. have constructed a similar baffle design for the GISMO instrument\cite{2008SPIE.7020E..66S,sharp2}.

\section{Stray Light and Spill}
\label{sec:stray_light}
 The frustum of rays defined by the FOV edges and the cleanly imaged entrance pupil on the VPM represents, to first order, the only path of light to the focal plane.  Neglecting diffraction effects, the only way for sources beyond the FOV to illuminate the focal plane would be diffusively scattering radiation into the light path.  This behavior was approximated by imagining the optical components and support structure scatter geometric rays with a Lambertian probability distribution peaking in the direction of the local surface normal.  Sources scattering significant power into the beam were eliminated, baffled or blackened. The most troubling surfaces was found to be the 4 K baffling between the lenses and the 1 K baffling between the focal plane and final lens. As a first effort to eliminate this effect, the walls are blackened with steelcast. Glint baffles, in the form of ribs, will surround the optical path to further eliminate large angular scattering.

Diffraction of light around the optical components is also a source of extra loading on the detectors. To protect against warm spill from Earth-shine, Moon-shine, Sun-shine or other strong sources of in-band radiation, the primary and secondary mirror must be oversized and a co-moving ground shield will surround the telescope. To predict the loading from diffraction beyond these elements, a series of POP calculations was performed in ZEMAX. As described in section \ref{sec:analysis}, a Gaussian beam representing the main beam of the horn is transformed through the system to predict the illumination of each surface.  Keeping track of the fraction of the beam lost at each surface and normalizing via $A\Omega = \lambda^2$ at the sky, the loading from spill beyond each surface can be calculated\cite{1966raas.book.....K}. Combining the POP results with an atmosphere model generated by ATM\footnote{ATM is authored and maintained by Juan Pardo.  The software is available at \mbox{/www.mrao.cam.ac.uk/~bn204/alma/atmomodel.html}} (assuming 1 mm precipitable water vapor and $45^\circ$ elevation angle), we find the expected detector loading is $\sim 2.2$ pW. Table \ref{tab:loading} breaks the sources of loading down into categories.   The total loading is dominated by the atmosphere.

\begin{table}[t]
\begin{center}
\caption{\label{tab:loading}Predicted total thermal load on the detector assuming a generic IR filtering scheme.}
\begin{tabular}{ll}
 Source of loading & Power (pW)\\
 \toprule
 CMB & 0.225\\
 Atmosphere & 1.34\\
 Warm spill & 0.136\\
 Cold spill & 0.19\\
 Telescope emission & 0.274 \\
 \midrule
 Total & 2.17\\
\end{tabular}
\end{center}
\end{table}

\section{Tolerances}
\label{sec:tol}

Monte Carlo (MC) simulations were performed to determine the tolerance for 144 degrees of freedom (DOF) of the optical design. These DOF included the shape, position, tip, tilt, thickness, material properties, and surface irregularity for all optical components. For each MC realization, all DOF were randomly perturbed around a nominal value according to a normal distribution with a specified width.  The secondary is then refocused via three steps of the DLS algorithm to minimize the WFE at the focal plane for each MC realization.  The WFE of all sampled fields was then recorded. 

Based upon 21,000 MC simulations, the tolerances for the optical design are extremely forgiving. The generous tolerances for the telescope are to be expected for this long wavelength band. The most challenging tolerance is setting the distance between the focal plane and nearest lens within 2 mm - owing to the relatively fast $f/2$ focal ratio at the focal plane.  All tolerances should be achievable by standard machining and assembly techniques.  For the MC simulations, 98\% of the simulated telescope instances still had Strehl ratios greater than 0.993 across the entire FOV. 

\section{Conclusions}
\label{sec:conclusions}

To study inflation, CLASS will use four telescopes at three frequencies to search for the B-mode signal in the CMB and thus infer the energy scale of the potential driving inflation. The 40 GHz channel of CLASS will be an essential component in the full multi-frequency CLASS instrument. To this end, we have developed a new telescope design for ground based observation of CMB polarization at 40 GHz with a front-end VPM modulator.  This new design separates the key tasks of imaging the cold stop onto the VPM and imaging the sky onto the focal plane by delegating the former task to a two mirror fore-optics section and the later to a two lens re-imaging cold optics section.  The telescope will cleanly map the horn antenna beams from the focal plane to $1.5^\circ$ FWHM beams on the sky. The peaks of the cross-polar response of the telescope are expected to be at least 40 dB lower than the main beam peak. A tolerance analysis indicates the presented design delivers robust performance for a wide range of perturbations. Design and construction of the VPM and optical components has begun, with full telescope integration to follow.  

\bibliography{OpticsBib}   

\begin{thebibliography}{10}

\bibitem{Guth}
A.~H. {Guth}, ``{Inflationary universe: A possible solution to the horizon and
  flatness problems},'' {\em Phys.~Rev.~D}~{\bf 23}, pp.~347--356, Jan. 1981.

\bibitem{Linde}
A.~D. {Linde}, ``{A new inflationary universe scenario: A possible solution of
  the horizon, flatness, homogeneity, isotropy and primordial monopole
  problems},'' {\em Physics Letters B}~{\bf 108}, pp.~389--393, Feb. 1982.

\bibitem{Albrecht}
A.~{Albrecht} and P.~J. {Steinhardt}, ``{Cosmology for grand unified theories
  with radiatively induced symmetry breaking},'' {\em Physical Review
  Letters}~{\bf 48}, pp.~1220--1223, Apr. 1982.

\bibitem{1982PhRvL..49.1110G}
A.~H. {Guth} and S.-Y. {Pi}, ``{Fluctuations in the new inflationary
  universe},'' {\em Physical Review Letters}~{\bf 49}, pp.~1110--1113, Oct.
  1982.

\bibitem{1983PhRvD..28..679B}
J.~M. {Bardeen}, P.~J. {Steinhardt}, and M.~S. {Turner}, ``{Spontaneous
  creation of almost scale-free density perturbations in an inflationary
  universe},'' {\em Phys.~Rev.~D}~{\bf 28}, pp.~679--693, Aug. 1983.

\bibitem{1997PhRvL..78.2054S}
U.~{Seljak} and M.~{Zaldarriaga}, ``{Signature of Gravity Waves in the
  Polarization of the Microwave Background},'' {\em Physical Review
  Letters}~{\bf 78}, pp.~2054--2057, Mar. 1997.

\bibitem{1997PhRvD..55.7368K}
M.~{Kamionkowski}, A.~{Kosowsky}, and A.~{Stebbins}, ``{Statistics of cosmic
  microwave background polarization},'' {\em Physical Review D}~{\bf 55},
  pp.~7368--7388, June 1997.

\bibitem{2009AIPC.1141...10B}
D.~{Baumann}, M.~G. {Jackson}, P.~{Adshead}, A.~{Amblard}, A.~{Ashoorioon},
  N.~{Bartolo}, R.~{Bean}, M.~{Beltr{\'a}n}, F.~{de Bernardis}, S.~{Bird},
  X.~{Chen}, D.~J.~H. {Chung}, L.~{Colombo}, A.~{Cooray}, P.~{Creminelli},
  S.~{Dodelson}, J.~{Dunkley}, C.~{Dvorkin}, R.~{Easther}, F.~{Finelli},
  R.~{Flauger}, M.~P. {Hertzberg}, K.~{Jones-Smith}, S.~{Kachru}, K.~{Kadota},
  J.~{Khoury}, W.~H. {Kinney}, E.~{Komatsu}, L.~M. {Krauss}, J.~{Lesgourgues},
  A.~{Liddle}, M.~{Liguori}, E.~{Lim}, A.~{Linde}, S.~{Matarrese}, H.~{Mathur},
  L.~{McAllister}, A.~{Melchiorri}, A.~{Nicolis}, L.~{Pagano}, H.~V. {Peiris},
  M.~{Peloso}, L.~{Pogosian}, E.~{Pierpaoli}, A.~{Riotto}, U.~{Seljak},
  L.~{Senatore}, S.~{Shandera}, E.~{Silverstein}, T.~{Smith}, P.~{Vaudrevange},
  L.~{Verde}, B.~{Wandelt}, D.~{Wands}, S.~{Watson}, M.~{Wyman}, A.~{Yadav},
  W.~{Valkenburg}, and M.~{Zaldarriaga}, ``{Probing Inflation with CMB
  Polarization},'' in {\em American Institute of Physics Conference Series},
  S.~{Dodelson}, D.~{Baumann}, A.~{Cooray}, J.~{Dunkley}, A.~{Fraisse}, M.~G.
  {Jackson}, A.~{Kogut}, L.~{Krauss}, M.~{Zaldarriaga}, and K.~{Smith}, eds.,
  {\em American Institute of Physics Conference Series} {\bf 1141},
  pp.~10--120, June 2009.

\bibitem{CMBforegrounds}
J.~{Dunkley}, A.~{Amblard}, C.~{Baccigalupi}, M.~{Betoule}, D.~{Chuss},
  A.~{Cooray}, J.~{Delabrouille}, C.~{Dickinson}, G.~{Dobler}, J.~{Dotson},
  H.~K. {Eriksen}, D.~{Finkbeiner}, D.~{Fixsen}, P.~{Fosalba}, A.~{Fraisse},
  C.~{Hirata}, A.~{Kogut}, J.~{Kristiansen}, C.~{Lawrence}, A.~M.
  {Magalha\~{}Es}, M.~A. {Miville-Deschenes}, S.~{Meyer}, A.~{Miller}, S.~K.
  {Naess}, L.~{Page}, H.~V. {Peiris}, N.~{Phillips}, E.~{Pierpaoli},
  G.~{Rocha}, J.~E. {Vaillancourt}, and L.~{Verde}, ``{Prospects for polarized
  foreground removal},'' in {\em American Institute of Physics Conference
  Series},  S.~{Dodelson}, D.~{Baumann}, A.~{Cooray}, J.~{Dunkley},
  A.~{Fraisse}, M.~G. {Jackson}, A.~{Kogut}, L.~{Krauss}, M.~{Zaldarriaga}, and
  K.~{Smith}, eds., {\em American Institute of Physics Conference Series} {\bf
  1141}, pp.~222--264, June 2009.

\bibitem{detectors}
K.~Rostem, C.~L. Bennett, D.~T. Chuss, N.~P. Costen, E.~Crowe, K.~Denis, J.~R.
  Eimer, N.~Louri, T.~Marriage, S.~H. Moselely, T.~R. Stevenson, D.~Towner,
  K.~U-Yen, G.~M. Voellmer, E.~J. Wollack, and L.~Zeng, ``Detector architecture
  of the cosmology large angular scale surveyor status and preliminary
  results,'' {\em Millimeter, Submillimeter, and Far-Infrared Detectors and
  Instrumentation for Astronomy VI}~{\bf 8452}, SPIE, 2012.

\bibitem{vpm_theory}
D.~T. {Chuss}, E.~J. {Wollack}, S.~H. {Moseley}, and G.~{Novak},
  ``Interferometric polarization control,'' {\em Applied Optics}~{\bf 45}, July
  2006.

\bibitem{vpm_example}
M.~{Krejny}, D.~{Chuss}, C.~D. {D'Aubigny}, D.~{Golish}, M.~{Houde}, H.~{Hui},
  C.~{Kulesa}, R.~F. {Loewenstein}, S.~H. {Moseley}, G.~{Novak}, G.~{Voellmer},
  C.~{Walker}, and E.~{Wollack}, ``The {H}ertz/{VPM} polarimeter: design and
  first light observations,'' {\em Applied Optics}~{\bf 47}, Aug. 2008.

\bibitem{2012ApOpt..51..197C}
D.~T. {Chuss}, E.~J. {Wollack}, R.~{Henry}, H.~{Hui}, A.~J. {Juarez},
  M.~{Krejny}, S.~H. {Moseley}, and G.~{Novak}, ``{Properties of a
  variable-delay polarization modulator},'' {\em Applied Optics}~{\bf 51},
  p.~197, Jan. 2012.

\bibitem{1987IJIMW...8.1165M}
J.~A. {Murphy}, ``{Distortion of a simple Gaussian beam on reflection from
  off-axis ellipsoidal mirrors},'' {\em International Journal of Infrared and
  Millimeter Waves}~{\bf 8}, pp.~1165--1187, Sept. 1987.

\bibitem{2006ApOpt..45.5168M}
C.~{Macculi}, M.~{Zannoni}, O.~A. {Peverini}, E.~{Carretti}, R.~{Tascone}, and
  S.~{Cortiglioni}, ``{Systematic effects induced by a flat isotropic
  dielectric slab},'' {\em Applied Optics}~{\bf 45}, pp.~5168--5184, July 2006.

\bibitem{afsar85}
M.~N. Afsar and K.~J. Button, ``{Millimeter-Wave Dielectric Measurement of
  Materials},'' {\em Proceedings of the IEEE}~{\bf 73}, pp.~1160--1166, january
  1985.

\bibitem{HDPE_contraction}
G.~Schwarz, ``Thermal expansion of polymers from 4.2 k to room temperature,''
  {\em Cryogenics}~{\bf 28}(4), pp.~248 -- 254, 1988.

\bibitem{Jackson}
J.~D. {Jackson}, {\em {Classical Electrodynamics, 3rd Edition}}, Wiley, July
  1998.

\bibitem{LZthesis}
L.~{Zeng}, {\em {Polarimetry in Astrophysics and Cosmology}}.
\newblock PhD thesis, The Johns Hopkins University, 2012.

\bibitem{2010ITAP...58.1383Z}
L.~{Zeng}, C.~L. {Bennett}, D.~T. {Chuss}, and E.~J. {Wollack}, ``{A Low
  Cross-Polarization Smooth-Walled Horn With Improved Bandwidth},'' {\em IEEE
  Transactions on Antennas and Propagation}~{\bf 58}, pp.~1383--1387, Apr.
  2010.

\bibitem{Xpol}
A.~C. {Ludwig}, ``{The definition of cross polarization.},'' {\em IEEE
  Transactions on Antennas and Propagation}~{\bf 21}, pp.~116--119, 1973.

\bibitem{grid}
G.~M. {Voellmer}, C.~{Bennett}, D.~T. {Chuss}, J.~{Eimer}, H.~{Hui}, S.~H.
  {Moseley}, G.~{Novak}, E.~J. {Wollack}, and L.~{Zeng}, ``{A large
  free-standing wire grid for microwave variable-delay polarization
  modulation},'' in {\em Society of Photo-Optical Instrumentation Engineers
  (SPIE) Conference Series},  {\em Society of Photo-Optical Instrumentation
  Engineers (SPIE) Conference Series} {\bf 7014}, Aug. 2008.

\bibitem{2011RScI...82h6101E}
J.~R. {Eimer}, C.~L. {Bennett}, D.~T. {Chuss}, and E.~J. {Wollack}, ``{Note:
  Vector reflectometry in a beam waveguide},'' {\em Review of Scientific
  Instruments}~{\bf 82}, p.~086101, Aug. 2011.

\bibitem{Ruze}
J.~{Ruze}, ``{Antenna Tolerance Theory -- A Review},'' {\em IEEE
  Proceedings}~{\bf 54}, pp.~633--642, Apr. 1966.

\bibitem{steelcast}
E.~J. {Wollack}, D.~J. {Fixsen}, R.~{Henry}, A.~{Kogut}, M.~{Limon}, and
  P.~{Mirel}, ``{Electromagnetic and Thermal Properties of a Conductively
  Loaded Epoxy},'' {\em International Journal of Infrared and Millimeter
  Waves}~{\bf 29}, pp.~51--61, Jan. 2008.

\bibitem{2008SPIE.7020E..66S}
E.~H. {Sharp}, D.~J. {Benford}, D.~J. {Fixsen}, S.~F. {Maher}, C.~T. {Marx},
  J.~G. {Staguhn}, and E.~J. {Wollack}, ``{Design and performance of a
  high-throughput cryogenic detector system},'' in {\em Society of
  Photo-Optical Instrumentation Engineers (SPIE) Conference Series},  {\em
  Society of Photo-Optical Instrumentation Engineers (SPIE) Conference Series}
  {\bf 7020}, Aug. 2008.

\bibitem{sharp2}
E.~H. {Sharp}, D.~J. {Benford}, D.~J. {Fixsen}, S.~H. {Moseley}, J.~G.
  {Staguhn}, and E.~J. {Wollack}, ``{Stray Light Suppression in the Goddard
  IRAM 2-Millimeter Observer (GISMO)},'' in {\em Society of Photo-Optical
  Instrumentation Engineers (SPIE) Conference Series},  {\em Society of
  Photo-Optical Instrumentation Engineers (SPIE) Conference Series} {\bf 8452},
  2012.

\bibitem{1966raas.book.....K}
J.~D. {Kraus}, {\em {Radio astronomy, 2nd Edition}}, Cygnus-Quasar Books, 1986.

\end{thebibliography}
\bibliographystyle{spiebib}   

\end{document}